# An insertable glucose sensor using a compact and cost-effective phosphorescence lifetime imager and machine learning

Artem Goncharov[1,2,3,§], Zoltan Gorocs[1,2,3,§], Ridhi Pradhan[4], Brian Ko[4], Ajmal Ajmal[5], Andres Rodriguez[5], David Baum[1], Marcell Veszpremi[1], Xilin Yang[1,2,3], Maxime Pindrys[6], Tianle Zheng[7], Oliver Wang[1], Jessica C. Ramella-Roman[5], Michael J. McShane[4,8], Aydogan Ozcan[1,2,3,9,*]

[1]Electrical & Computer Engineering Department, [2]Bioengineering Department, [3]California NanoSystems Institute (CNSI), University of California, Los Angeles, CA 90095 USA, [4]Department of Biomedical Engineering, Texas A&M University, College Station, TX 77843 USA, [5]Department of Biomedical Engineering, Florida International University, FL 33199 USA, [6]Department of Physics, University of Connecticut, Storrs, CT 06269 USA, [7]Department of Computer Science, University of California, Los Angeles, CA 90095 USA, [8]Department of Materials Science and Engineering, Texas A&M University, College Station, TX 77843 USA, [9]Department of Surgery, University of California, Los Angeles, CA 90095 USA

§ Equal contribution

*Corresponding Author: ozcan@ucla.edu

## Abstract

Optical continuous glucose monitoring (CGM) systems are emerging for personalized glucose management owing to their lower cost and prolonged durability compared to conventional electrochemical CGMs. Here, we report a computational CGM system, which integrates a biocompatible phosphorescence-based insertable biosensor and a custom-designed phosphorescence lifetime imager (PLI). This compact and cost-effective PLI is designed to capture phosphorescence lifetime images of an insertable sensor through the skin, where the lifetime of the emitted phosphorescence signal is modulated by the local concentration of glucose. Because this phosphorescence signal has a very long lifetime compared to tissue autofluorescence or excitation leakage processes, it completely bypasses these noise sources by measuring the sensor emission over several tens of microseconds after the excitation light is turned off. The lifetime images acquired through the skin are processed by neural network-based models for misalignment-tolerant inference of glucose levels, accurately revealing normal, low (hypoglycemia) and high (hyperglycemia) concentration ranges. Using a 1-mm thick skin phantom mimicking the optical properties of human skin, we performed *in vitro* testing of the PLI using glucose-spiked samples, yielding 88.8% inference accuracy, also showing resilience to random and unknown misalignments within a lateral distance of ~4.7 mm with respect to the position of the insertable sensor underneath the skin phantom. Furthermore, the PLI accurately identified larger lateral misalignments beyond 5 mm, prompting user intervention for re-alignment. The misalignment-resilient glucose concentration inference capability of this compact and cost-effective phosphorescence lifetime imager makes it an appealing wearable diagnostics tool for real-time tracking of glucose and other biomarkers.

## Introduction

Wearable sensors have emerged as a minimally invasive monitoring platform for real-time detection of key physiological and healthcare biomarkers, including blood pressure[1,2], oxygen level[2], temperature[3,5,6], pH[3,4], glucose level[4,7-9], electrolytes[10-12], hormones[13-14], and metabolites[4,15]. Recent advances in hardware miniaturization[16], material science[17] and data analysis techniques[18] have allowed for robust *in vivo* sensing technologies using various samples such as sweat[19], saliva[20,21] and interstitial fluid (ISF)[22,23]. Continuous glucose monitoring (CGM) devices stand out as an important class of wearable sensors utilized for real-time tracking of glucose levels[24-26]. Glucose is a biomarker for diabetes[27], a chronic metabolic disorder, affecting over 450 million individuals globally, with projected patient numbers expected to increase to over 550 million by 2030[28]. Daily monitoring of glucose levels is required for the proper management of diabetes to prevent hypo- and hyperglycemic events, which may cause severe health complications, such as cardiovascular disease (CVD), kidney failure, and neuropathy[29]. Conventional glucose monitoring methods rely on highly invasive and laborious finger pricking procedures[30,31]; CGM systems have evolved as a user-friendly alternative approach, facilitating real-time tracking of patients' glucose levels.

Typical CGMs employ subcutaneous electrodes connected to a wearable transmitter to collect glucose information from ISF through enzymatic electrochemical reactions[25]. Several CGMs based on this technology have received approvals from the Food and Drug Administration (FDA) for human use[32-34]. However, despite their widespread adoption, these CGMs are associated with relatively high costs (i.e., ~$450/month) due to the short lifetime of the sensing electrodes (~1-2 weeks) and cause patient discomfort during the electrode implantation and operation[25]. In addition, the insertion of the electrodes under the skin may damage surrounding cells and potentially cause tissue inflammation, leading to skin irritation and reduced testing accuracy[35-36].

To address some of these challenges, alternative methods based on optical signal transduction have been proposed[24,37]. In particular, fluorescence[24,38,39], colorimetric detection[40], near-infrared spectroscopy[41,42], and surface-enhanced Raman spectroscopy (SERS)[43], are some of the emerging optical modalities for noninvasive and minimally invasive monitoring of glucose levels. Among these methods, fluorescence-based minimally invasive CGMs have been successfully tested *in vivo* and, to date, represent the only FDA-approved optical CGM[24] platform. This CGM system comprises an implantable glucose-specific fluorescent sensor, which measures glucose concentration in ISF and sends the data to a removable wearable transmitter, connected to a smartphone for data display. The sensor is implanted subcutaneously into the upper arm and can operate for up to 90 days post-implantation, offering a longer-term and more cost-effective solution (i.e., ~$120/month) compared to electrochemical CGMs[25]. However, this platform has lower testing accuracy and requires additional calibration through finger pricking ~2 times per day, hindering the minimally invasive nature of the technology[25]. In addition, the relatively large size of the implant (i.e., 3.3 mm x 15 mm) may lead to tissue inflammation, causing patient discomfort and potentially reducing sensor accuracy[44,45]. Therefore, further advancements in optical CGMs are necessary to enhance the usability, stability, and overall robustness of these devices for their widespread adoption in POC settings.

As an alternative to existing optical sensor technologies, phosphorescence is an appealing optical technique for CGM systems due to a substantially longer lifetime of phosphorescence emission (>10 µs) compared to human tissue autofluorescence (≤10 ns), allowing for a superior temporal separation between native fluorophores and phosphorescent sensing signals. Consequently, phosphorescence-based sensors provide a better signal-to-noise ratio (SNR) in comparison with fluorescence-based designs, leading to a higher signal transduction efficiency[46,47]. Palladium and platinum porphyrins form a major class of oxygen-sensitive phosphorescent probes reported in the literature for biosensing applications[48,49]. Both the emission intensities and lifetimes of these phosphors vary in response to differences in the local oxygen concentrations, and with the addition of enzymes consuming the analyte of interest, phosphor emissions can become sensitive to the analyte concentrations. Therefore, phosphorescence-based sensors utilizing this "indirect" sensing approach have been developed for continuous glucose monitoring in ISF[50-52]. These sensors are typically smaller in size (i.e., 1.2 mm x 6.5 mm) compared to the conventional fluorescence-based sensors, enabling improved biocompatibility and easier subcutaneous insertion[52].

With these advances in phosphorescence-based biosensor fabrication, the development of appropriate readout hardware that is both compact/wearable and cost-effective becomes essential for accurate and real-time inference of glucose levels through the skin[53-55]. Conventional wearable phosphorescence readers deliver excitation light to phosphors and use a *single* photodetector to capture the signal intensity and/or lifetime of phosphorescent emissions[55,56]. However, this single pixel-based readout approach encounters challenges in uncontrolled real-life testing environments, mostly related to reader misalignments due to patient movement. These misalignments can significantly alter the effective signal captured by the photodetector, leading to high variability in the output glucose levels. Some optical CGMs overcome this issue by integrating both the sensing assay and the miniaturized readout module within the implanted sensor, utilizing an external wearable reader solely for data transfer and display[24]. A major drawback of this method is the relatively large size of the sensor (i.e., 3.3 mm x 15 mm), which lowers its biocompatibility and requires surgical operation for implantation and removal, thereby limiting its usability[25]. Furthermore, owing to its large size, this sensor causes inflammation in the surrounding tissue, leading to variability in glucose diffusion rate and necessitating invasive re-calibration. In contrast, a *passive* phosphorescence-based insertable sensor that solely contains the sensing assay would be significantly more compact, allowing for easier insertion with less tissue inflammation; however, it would necessitate an external wearable and cost-effective hardware for the signal readout. Therefore, efficient alignment control of the external reader positioning is particularly important to ensure accurate sensor readout through the skin and minimize misalignment-induced glucose concentration inference errors. Single photodetector-based readers, in general, lack spatial information about the relative sensor location, preventing effective control over reader alignment. Imaging-based readout configurations present a promising alternative owing to their ability to capture emitted signals in a highly multiplexed manner by capturing sensor images[57]. Such imaging-based solutions can provide spatial information about the sensor's location, enabling alignment control of the reader.

Here, we introduce a compact and cost-effective phosphorescence lifetime imager (PLI) designed for misalignment-tolerant inference of glucose levels using an insertable passive hydrogel sensor (Figure 1). The PLI continuously captures both phosphorescence intensity and lifetime images of an insertable sensor through the skin and automatically processes the acquired image data using trained neural network models to (1) assess the alignment of the PLI's field-of-view (FOV) with respect to the implanted sensor location and inform the user in case of misalignments, and (2) classify the glucose concentration levels as low (<70 mg/dL), normal (70-125 mg/dL), and high (>125 mg/dL); see Figure 1d.[58-60]. We note that glucose concentrations in the 125-180 mg/dL range are associated with prediabetes, another widespread health condition linked to a higher risk of developing type 2 diabetes and CVD. Just in the US, over 90 million patients are affected by this disease, with 80% being unaware of their condition[61]. Hence, early detection of both diabetes and prediabetes through CGM systems holds promise to lower diabetes prevalence and improve the quality of glucose management.

We tested the performance of our PLI system *in vitro* using glucose-spiked deionized (DI) water samples using a 1-mm thick skin phantom to simulate the optical properties of the skin tissue. Our results demonstrated consistent sensor response for glucose concentrations in the range of 0-250 mg/dL and a good inter-sensor repeatability with a coefficient of variation (CV) of <15% regardless of the glucose level. In addition, we integrated our PLI with a neural network-based framework for misalignment resilient inference of glucose levels and showed an accuracy of 88.8% for the classification of glucose concentration levels (low/normal/high) while being resilient to random/unknown misalignments within a lateral distance of ~4.7 mm. Furthermore, this compact and cost-effective PLI reader achieved 100% accuracy in autonomously identifying larger physical misalignments beyond 5 mm, prompting the users to re-align the reader and minimizing misalignment-induced glucose level measurement inaccuracies.

This misalignment-resilient glucose level inference capability of the PLI system, coupled with the biocompatibility and small size of the insertable phosphorescence biosensor, make our platform an attractive CGM system for personalized monitoring of ISF glucose levels. The same imaging platform also permits multiplexed sensing capabilities to implement parallel monitoring of multiple biomarkers through the skin, thereby expanding the application scope of this technology beyond glucose monitoring, potentially making it a versatile wearable sensing platform.

# Results and Discussion

*Design of the insertable phosphorescence-based glucose sensor*

The insertable phosphorescence-based sensor was fabricated from a hydrogel (PEGDA) and had a "barcode" structure, featuring four spatially separated discrete compartments filled with glucose-sensitive *test channels* and glucose-insensitive *control channels* positioned in duplicates[52] (Figure 1b, see the *Fabrication of insertable/implantable glucose sensors* section in Methods). These assays utilize palladium-based phosphorescent dyes (i.e., palladium benzoporphyrins [PdBP]) with typical emission lifetimes in the 50-250 μs range, three orders of magnitude longer than the maximal autofluorescence lifetime of endogenous fluorophores like melanin and collagen, which do not exceed 10 ns. This enables superior temporal separation of the sensor responses from the native fluorescence, leading to better SNR of these phosphorescence-based sensors compared to other luminescence-based designs, and therefore allowing for deeper sensor operation from subcutaneous tissue layers where ISF is most abundant.

In addition to the phosphor dyes, test channel assays also contain glucose-specific enzymes (i.e., glucose oxidase and catalase), making phosphorescence intensities and lifetimes affected by the local changes in the glucose level. In contrast, the control channel assays only contain phosphors and provide constant signals, regardless of the glucose concentration in the local environment. The insertable sensors demonstrated stable phosphorescence lifetime response for up to 12 weeks, with enzyme activity maintained at over >80% for up to 4 weeks[52]. In addition, it showed good biocompatibility for up to 7 months in real-world environments within the animal (i.e., pig) body[52]. The combination of high SNR, small footprint and biocompatibility renders these insertable phosphorescence-based sensors a promising signal transduction probe for next-generation CGM systems.

*Design of the phosphorescence lifetime imager (PLI) and image processing pipeline*

Along with the development of phosphorescence-based insertable biosensors, appropriate readout hardware is also needed for a robust glucose concentration inference with resilience against potential misalignments of the wearable reader under real-life operating conditions. Toward this goal, we created a compact and cost-effective PLI, which captures phosphorescence intensity and lifetime images of an insertable/implantable glucose sensor and utilizes these images for misalignment-resilient inference of glucose levels through the skin. The PLI reader has a compact footprint (~5 cm x 5 cm) and comprises an imaging system with a red excitation LED (633 nm) operating in a pulse mode and optical filters, accommodating the spectral properties of the phosphors (630 nm excitation/810 nm emission) (Figure 2b). The imaging system has a sensing FOV of ~9.6 mm x 7.2 mm, allowing the capture of the whole sensor image. Excitation intensity was adjusted to ~10-fold below the American National Standards Institute (ANSI) safety exposure limit for safe operation[62]. For each measurement, the PLI captured 10 successive phosphorescence images, utilized to generate the phosphorescence intensity and lifetime images of the sensing FOV, which were further processed by a neural network-based framework for misalignment-resilient glucose level inference with a repetition rate of ~1 min (see *PLI design* and *PLI operation* sections in Methods for details). For each phosphorescence lifetime image, the reader acquires 11 timelapse phosphorescence images, evenly distributed within 200 µs time interval, starting immediately after the excitation LED turns off, to capture the decay of the sensor emission through the skin; see Figures 1d-2c. These 11 timelapse images are processed by an image analysis algorithm to generate the phosphorescence intensity and lifetime images of the sensing FOV, as detailed in the Methods section (see the *Generation of phosphorescence intensity and lifetime images* in Methods). Lifetime responses of each implanted sensor are derived by averaging the pixels within rectangular masks superimposed on each of the 4 sensor channels (2 test channels and 2 control channels), yielding 4 lifetime values per implanted sensor barcode behind the skin phantom. Typical lifetime response values for the test channels ranged from ~80 µs for DI water (i.e., 0 mg/dL glucose concentration) to ~170 µs for high glucose concentration levels (≥150 mg/dL) and stayed around 180-190 µs for the control channels. Details of the *in vitro* testing of insertable glucose sensor responses to different glucose concentrations are reported in the next section.

*In vitro testing of insertable glucose sensors quantified using PLI*

We characterized the *in vitro* performance of the insertable sensors and our custom-designed PLI using glucose-spiked DI water samples (see the *In vitro data collection procedures* section in Methods); for these experiments, we utilized a skin phantom (1 mm, Type 1-2, see the *Skin phantom fabrication* section in Methods) to simulate the optical properties of a bulk skin tissue (Figure 2a). Leveraging the imaging-based design of the PLI system, we measured sensor responses through the skin phantom at different locations within the reader FOV. For the analysis of the misalignment tolerance of the system, we defined two regions of interest (ROIs), namely *aligned* and *misaligned*. The aligned region was selected as a 3.2 mm x 3.4 mm rectangle centered around the center of the reader FOV, while the misaligned region was set as the remaining area, outside of the aligned region (Figure 3a).

In an ideal PLI operation scenario, the insertable biosensor would maintain a fixed position at the center of the reader FOV (i.e., the *ideal* location), exhibiting stable excitation power and constant imaging condition. However, real-life operation scenarios may involve random wearable reader misalignments, leading to fluctuations in the effective sensor excitation profile and variability in the sensor positioning within the reader FOV, which may alter the captured sensor measurements. In this context, the aligned region represents an acceptable *misalignment-tolerant* zone: as long as the sensor remains within this zone, the PLI captures sensor measurements for accurate inference of glucose levels despite random deviations from the ideal location (i.e., FOV center). Otherwise, if the sensor falls outside of the aligned region, i.e., into the misaligned area, using a trained neural network, the PLI reader detects it and prompts the user for re-alignment to mitigate potential readout variabilities induced by large misalignments (detailed in the next subsection).

Prior to the development of the neural network-based analysis for glucose level inference, we first tested the responses of the insertable sensor through a 1-mm thick skin phantom at different locations within the PLI reader FOV. Figure 3b depicts sensor lifetime responses from 3 representative locations within the reader FOV, including 1 ideal location (i.e., the center of the aligned region), 1 non-ideal location (i.e., at the border of the aligned region) and 1 misaligned location within the misaligned region. The phosphorescence lifetime responses of the test channels at both the ideal and non-ideal locations within the aligned zone exhibit a strong correlation with glucose concentrations in the 0-150 mg/dL range. In the meantime, the control channels at both locations demonstrate stable responses, independent of the glucose concentration as expected, with less than 5% CV (i.e., intra-sensor CV), validating consistent sensor operation (Figures 3b(i,ii), S1; Video S1). In addition, inter-sensor variability, assessed between the lifetime responses of 8 ($N_{sens}$) tested implantable sensors, shows less than 15% CV (i.e., inter-sensor CV) for all three glucose levels (low/normal/high), indicating a repeatable PLI operation within the aligned region. The comparable performances of the sensor responses from both the ideal and non-ideal locations within the aligned region further justify our definition of the alignment zone.

In contrast to the aligned region, only a single test channel (Test 1) from the misaligned sensor shows a strong correlation with the spiked glucose concentration levels (Figure 3b(iii)). The second test channel (Test 2) behaves similarly to the control channels, providing a constant response regardless of the glucose level. This undesired behavior can be attributed to the low SNR of this misaligned test channel and the interference of the emissions from the control channels, scattered by the skin phantom layer. Additionally, inter-sensor repeatability for the misaligned location is lower compared to aligned locations, with inter-sensor CV exceeding 15%. This inferior performance of the insertable sensor from the misaligned location underscores the need for alignment control for our wearable PLI to achieve robust inference of glucose levels.

Both the lifetimes and intensities of the phosphorescence emission are influenced by the local changes in the glucose concentration. Therefore, in addition to looking at the lifetime responses, we also explored the intensity responses as an alternative detection modality. We observed that lifetime signals exhibit significantly lower inter-sensor variability compared to the intensity signals, making phosphorescence lifetime the preferred modality for consistent and accurate inference of glucose levels through the skin. For instance, for the DI water samples (i.e., 0 mg/dL glucose), the inter-sensor CV for the intensity responses exceeded 25% and 15% for the test and control channels, respectively, while the CVs of the lifetime responses remained under 11% for all the channels (Figure S2). Moreover, the average CVs of the intensity responses to a glucose concentration sweep for the locations depicted in Figure 3b exceeded 50% for all three locations, including the aligned cases (i.e., ideal and non-ideal) (Figure S3). This observed higher variability in the phosphorescence intensity responses may originate from the manual fabrication of the insertable sensors or variations in the ambient environment (i.e., the oxygen concentration in DI water samples), which primarily affect phosphorescence intensities rather than lifetimes.

We also tested the linearity and repeatability of the sensor responses under different levels of pixel binning. Captured phosphorescence lifetime images without pixel binning consisted of 800x600 pixels with a pixel size of 9 μm, resulting in ~110x90 pixels per channel of each insertable sensor. We used a baseline pixel binning size of N=8 (i.e., ~14x11 pixels per channel) for the *in vitro* testing results reported in our work. However, we also investigated sensor responses for much larger pixel binning sizes and observed a consistent sensor performance with < 15% inter-sensor CV up to N=30, which corresponds to ~4x3 pixels per test channel (Figure S4). Therefore, lower-resolution imaging systems with ~26x20 pixels and larger photodetector sizes (i.e., 270 μm) could potentially be utilized for the PLI design.

As illustrated in Figure 3b, the insertable sensors showed a good linearity up to a glucose concentration of ~150 mg/dL; however, for higher glucose concentrations, the response was saturated, likely indicating an enzyme-limited behavior of the sensors. A higher glucose diffusion rate at higher glucose concentrations might be saturating the enzyme activity; therefore, increasing the enzyme concentration may help to extend the linear range of our glucose sensors. Alternatively, employing a larger number of diffusion control crosslinked polyelectrolyte layers (i.e., CX layers) can slow down the glucose diffusion into the sensor matrix, consequently improving the sensor linearity; see the *Fabrication of insertable/implantable glucose sensors* sub-section of the Methods for details. Due to a tradeoff between the number of CX layers and the oxygen concentration in the environment, we did not increase the number of CX layers beyond 7 to maintain optimal testing conditions for the ambient oxygen environment (~21%) used in this study (see Figures S5-S6). Nevertheless, for testing at lower oxygen environments (e.g., ~5%), increasing the number of CX layers can yield a more linear sensor response across a larger glucose concentration range[52].

*Neural network-based analysis of PLI data for inferring glucose concentration levels*

Our PLI system used deep learning and convolutional neural network (CNN) models for the misalignment-resilient inference of glucose concentration levels from the phosphorescence intensity and phosphorescence lifetime images. The neural network-based analysis consisted of two separate CNN models: the alignment model (i.e., $CNN_{Alignment}$), which was used to assess the alignment of the PLI reader with respect to the position of the sensor underneath the skin phantom, and the classification model (i.e., $CNN_{Class}$) utilized to classify glucose concentrations between low, normal, and high ranges (Figure 4a). $CNN_{Alignment}$ automatically assessed the alignment of the PLI reader location by analyzing phosphorescence intensity images and identifying the sensor location within the image. If $CNN_{Alignment}$ located the sensor within the aligned region (i.e., 3.2 mm x 3.4 mm, defining the misalignment-tolerant zone), it predicted that the reader was properly aligned; conversely, if the sensor location fell into the misaligned region, the network predicted misalignment of the reader. In misaligned cases, the reader prompted the user for re-alignment, and the $CNN_{Class}$ was not used for such measurements. If and only if $CNN_{Alignment}$ identified that the reader was properly aligned, the measured sample was further processed by the $CNN_{Class}$, which utilized phosphorescence lifetime image measurements for glucose level classification. Based on this information processing pipeline, neural network-based glucose level inference accommodated up to 4.7 mm reader misalignment.

These alignment and classification network models were trained and optimized separately using a total of 680 samples comprising 200 aligned and 480 misaligned samples from 4 different insertable sensors, each tested on 10 glucose concentrations in the range of 0-250 mg/dL. To generate a diverse dataset representing various reader alignment scenarios, each glucose concentration was measured at 17 different sensor locations within the PLI reader FOV, including 5 aligned (1 ideal + 4 non-ideal) and 12 misaligned locations. The $CNN_{Alignment}$ was trained and optimized using all these 680 measurements, achieving 100% classification accuracy on the validation set. $CNN_{Alignment}$ accurately identified aligned and misaligned locations, benefitting from the spatial features present in the phosphorescence intensity images. At the same time, $CNN_{Class}$ was solely trained on 200 aligned samples since it would only be used if $CNN_{Alignment}$ identified that the reader was properly aligned; the optimal $CNN_{Class}$ model exhibited 89.3% accuracy on the validation set for the glucose level classification, including an accuracy of 93.7% and 87.5% for the ideal and non-ideal sensor locations, respectively (see the *Neural network-based analysis of PLI data* in the Methods section for details).

After this training stage, the optimized models were blindly tested on 672 new measurements from 4 additional implantable sensors, never used during the training stage. $CNN_{Alignment}$ showed the same 100% accuracy on the blind samples, correctly identifying 198 aligned and 472 misaligned measurements (Figure 4b). For all measurements identified as misaligned by $CNN_{Alignment}$, re-alignment of PLI would be needed, and $CNN_{Class}$ would not be applied to any of these samples. Consequently, only the measurements falling within the aligned region (i.e., 3.2 mm x 3.4 mm, misalignment-tolerant zone), as determined by the $CNN_{Alignment}$, were further processed by the $CNN_{Class}$. The performance of $CNN_{Class}$ on these samples showed an overall accuracy of 88.8% for the classification of low, normal, and high glucose levels, revealing 92.5% accuracy for the ideal sensor locations and 87.3% accuracy for the non-ideal sensor locations, as illustrated by confusion matrices in Figure 4c.

To further explore the implications of some of the false predictions by $CNN_{Class}$, we analyzed these confusion matrices and categorized false predictions into three classes based on the severity of their impact. In the context of the false model predictions on patient health, a misclassification between hypoglycemic (low glucose level) and hyperglycemic (high glucose level) events represents a severe error, leading to opposite patient treatment and exacerbating the patient's condition. Fortunately, our $CNN_{Class}$ did ***not*** confuse hypoglycemic and hyperglycemic events for any of the blinded test samples, as illustrated by the grey-colored cells of the confusion matrix. The second most critical error involves the misclassification of low or high glucose levels as normal, leaving abnormal glucose levels undetected. These cases are represented in the confusion matrices with the red-colored cells. The optimal $CNN_{Class}$ model showed 93.1% sensitivity (i.e., 54/58 true positives) for detecting low glucose levels and 96.7% sensitivity (i.e., 58/60 true positives) for detecting high glucose levels, which confirms the competitive performance of our PLI reader. Finally, the misclassification of normal glucose levels as low or high is another type of error. These predictions are depicted with yellow-colored cells in the confusion matrices. During the model optimization, we specifically prioritized $CNN_{Class}$ models with lower error rates for the two more severe error cases (grey and red colored cells in the confusion matrices) in order to minimize the negative impact of false predictions from our PLI reader.

The presented CNN models benefit from the collective behavior of the highly multiplexed signals at the captured phosphorescence images to achieve competitive accuracy for glucose level inference despite physical misalignments of the PLI reader. To better assess the impact of multiplexing on the model performance, we compared glucose level classification accuracy using images with lower resolution, controlled by the pixel binning size (N). $CNN_{Class}$ demonstrated accuracies of 84.8% and 83.8% when increasing N to 20 (40x30 pixels per sensor image) and 50 (16x12 pixels per image), respectively (Figure S7); $CNN_{Alignment}$ maintained 100% accuracy in both cases. These glucose concentration classification accuracies for large N suggest that lower-resolution imaging systems can be employed to reduce the cost of the PLI reader without a significant impact on its performance. These findings are also in line with our earlier results reported in Figure S4, demonstrating a consistent sensor performance with < 15% inter-sensor CV up to N=30 pixel binning.

The size of the aligned region creates a tradeoff between the usability of the reader and the accuracy of glucose testing. A smaller aligned region may increase testing accuracy. However, it requires a more precise reader positioning, leading to tedious manual alignment and reduced usability of the device. At the same time, large misalignments, particularly near the edge of the reader FOV, introduce additional excitation non-uniformities and aberrations, resulting in higher variabilities in captured sensor measurements and lower glucose testing accuracy. Evaluation of the same CNN models on locations beyond the baseline aligned zone outlined in Figure 3a, revealed that the $CNN_{Class}$ accuracy did not exceed 70% for such regions, with the highest error rate occurring for locations outside of the aligned zone (see Figure S8). Therefore, the 3.2 mm x 3.4 mm aligned zone accommodating up to ~5 mm lateral misalignments presents a balanced compromise between the PLI reader's ease-of-use and glucose testing performance, with a glucose concentration classification accuracy of 88.8%. In the future, the size of the aligned region can be extended by further increasing our imaging FOV and including more samples from different locations within the misaligned region in the training of the CNN models.

Finally, we should note that further testing of our platform is required to evaluate its feasibility for real-life operating conditions, which include low ambient oxygen environments (i.e., ~5% in the human body) and human body temperature (i.e., ~37 °C). For these future studies, we will also employ optimal sensor designs aimed at increasing the sensor linear operation range to cover higher glucose concentrations[52,56]. In addition, the usability of our technology can be improved by further miniaturization of the PLI reader to a watch-size footprint, which can be achieved by utilizing lens-less imaging designs[63,64] or miniaturized systems[65]. Finally, one can benefit from the multiplexing capabilities of the PLI reader and the multiplexed "barcode" structure of the insertable biosensors to extend our platform to parallel monitoring of different analytes, making it a versatile wearable sensor for various biomedical sensing/monitoring applications.

## Conclusions

In summary, we developed a CGM system based on insertable phosphorescence-based sensors paired with a compact and cost-effective phosphorescence lifetime imaging-based reader (PLI) for robust measurement of glucose levels with resilience to random reader misalignments. *In vitro* testing of our CGM platform using a 1-mm thick skin phantom revealed 88.8% accuracy for the classification of glucose levels covering low, normal, and high concentration ranges accommodating up to 4.7 mm misalignment. Furthermore, our PLI reader exhibited 100% accuracy in identifying larger misalignments beyond the misalignment-tolerant region, prompting user intervention for reader re-alignment, and ensuring appropriate reader positioning. We believe that the misalignment resilient glucose level inference capability of our PLI reader operating through the skin, coupled with the small size and biocompatibility of our insertable sensors, make this approach an appealing candidate platform for continuous ISF-based glucose monitoring.

## Methods

### *Fabrication of glucose-sensing alginate microparticles*

Phosphorescent signals from glucose sensing alginate microparticles were generated by oxygen-sensitive ethyl cellulose nanoparticles (ECNP) containing palladium (II) meso-tetra(4-carboxyphenyl)tetrabenzo-porphyrin) (Frontier Specialty Chemicals) phosphorescent dye (PdBP). PdBP-ECNPs were fabricated using the previously developed nano-emulsion method[52]. Glucose specificity in alginate microparticles was enabled by encapsulating PdBP-ECNP with glucose oxidase (GOx, Tokyo Chemical Industries) and catalase (Cat, Sigma-Aldrich, Inc.) using an emulsion technique. In brief, a 4% (w/v) sodium alginate aqueous solution (3.75 mL) was mixed with PdBP-ECNP (1.25 mL) suspension for 30 minutes by a nutating mixer. Separately, GOx (58.5 mg) and Cat (54.9 mg) were dissolved in 50 mM TRIS buffer (2.5 mL, pH 7.2, Sigma-Aldrich) by nutation. Further, the alginate mixture and the enzyme solution were mixed, generating a precursor solution. This solution was dropwise added and emulsified in a mixture of isooctane (10.8 mL, Avantor performance materials) with SPAN 85 (322 µL) using a homogenizer (8000 rpm). Further, isooctane (1.5 mL) and TWEEN 85 (175 µL) were added to the above mixture and stirred at the same speed for 15 seconds. During the last 50 seconds of the emulsification process, 10% (w/v) $CaCl_2$ (Sigma-Aldrich) solution (4 mL) was added to induce external gelation of the alginate microparticles. Next, the emulsion was transferred to a round bottom flask and gently stirred in a magnetic stirrer for 20 minutes. The microparticles were then centrifuged at 2000 g for two minutes and washed twice with DI water. Finally, the microparticles were coated with surface nanofilms of polyelectrolytes deposited using the layer-by-layer (LbL) assembly. The nanofilms consisted of crosslinked poly(allylamine hydrochloride) (PAH, Sigma-Aldrich) and poly(sodium-4-styrenesulfonate) (PSS, Sigma-Aldrich) polyelectrolyte bilayers (i.e., CX layers) and were designed for controlled diffusion of glucose inside the microparticles. The number of CX layers was optimized to 7 for the optimal sensor operation at the ambient oxygen environment (i.e., ~21%) used in this study (Figure S6). The number of CX layers can be increased for testing at low-oxygen environments, which will reduce the glucose diffusion rate and increase the linear range of sensor responses[52].

*Fabrication of insertable/implantable glucose sensors*

The hydrogel sensor was fabricated from PEGDA (i.e., polyethylene glycol diacrylate, Alfa Aesar) hydrogel using a previously reported soft lithography process[52]. In short, first, the 20% (w/v) PEGDA was mixed with 2% (v/v) photo-initiator solution and dispensed into a PDMS bottom mold. Further, a PDMS top mold with four discrete compartments was added squeezing the hydrogel solution between the PDMS top and bottom molds. Next, the squeezed hydrogel solution was placed under a UV lamp (360 nm, 10-15 mW/cm²) for 5 min to crosslink PEGDA monomers. The formed hydrogel was peeled off from PDMS molds and rinsed with DI water. PDMS top and bottom molds were fabricated by replica molding from 3D-printed master molds.

Two out of the four hydrogel compartments were pipetted with test channels (0.64 µL each) and two others with control channels (0.64 µL each). Test channel solution was prepared by mixing glucose-sensing alginate microparticles (see the *Fabrication of glucose-sensing alginate microparticles* section) with the hydrogel solution. The control channel solution was prepared by mixing oxygen-insensitive nanoparticles with the hydrogel solution. Oxygen-insensitive nanoparticles were fabricated by encapsulating palladium(II) tetramethacrylated benzoporphyrin (PdBMAPP, PROFUSA) within poly(vinylidene chloride-co-acrylonitrile) hydrogel, as previously described[54]. The hydrogel with pipetted test and control channels was polymerized under a UV lamp for 5 minutes. The total cost for one insertable sensor was < $0.3 (Table S1).

*LED driver circuit*

A custom LED driver circuit was designed to operate the excitation LED in a pulse mode (Figure S9). The circuit is based on a 1 A constant-current driver (Analog Devices Inc.) which contains internal pulse width modulation (PWM) circuitry allowing for pulsed operation with custom pulse on/off times. The operation of the PMW pin was set by a timer circuitry consisting of an 8 MHz oscillator (Analog Devices Inc.), a 14-stage shift register (Onsemi), and a quadruple NAND gate (Texas Instruments). The oscillator generated square pulses with 125 ns (50% duty cycle) baseline period, and the shift register further divided baseline oscillation frequency, yielding square pulses with 2,000 µs and 500 µs periods. Finally, a quadruple NAND gate combined these two square pulses to generate the desired waveform with 250 µs ($t_{on}$) and 1000 µs ($t_{off}$) on and off times, respectively. The waveform shape can be adjusted to accommodate different lifetime ranges by changing the baseline oscillation frequency or the output pins of the shift register. The constant-current driver was powered by an external supercapacitor to supply the required input voltage (i.e., $V_{In}$ = 6V). The LED was connected to GND through an N-channel MOSFET (Vishay Siliconix), enabling fast switching between on/off cycles. The LED driver circuit was powered by the output supply voltage pin (5 V, 100 mA) of the PLI camera. The circuit was designed using Altium circuit design software.

The excitation time (i.e., $t_{excitation}$), determined as the total LED activation time used for excitation of the phosphors to generate timelapse phosphorescence images, was optimized to $t_{excitation}$ = 70 ms (Figure S10). During this time, the LED was operated in a pulse mode, and the reader periodically captured phosphorescent responses, accumulating emissions over multiple excitation cycles, to achieve reliable SNR. Shorter excitation times resulted in a low SNR of the captured phosphorescence intensities, while $t_{excitation}$ > 70 ms depleted the voltage at the constant-current driver, leading to fluctuations in the LED output power.

*PLI operation*

A custom-designed graphical user interface (GUI) was developed to control the operation of the PLI (Figure S11). GUI contained input fields to specify the operation parameters of the PLI camera system, including shutter parameters, exposure time and gain. For each successive phosphorescence lifetime image measurement, the camera shutter was programmed to capture up to 11 timelapse phosphorescence images and the background image after the LED on cycle. For each of these images, the camera shutter was activated periodically, and the user could specify both the shutter activation times and the shutter on/off time periods. The shutter activation times for the 11 images were adjusted to cover a 200 µs time interval after the LED on cycle with a 20 µs gap between subsequent activations, i.e., $t_1 = 0$ µs, $t_2 = 20$ µs, $t_3 = 40$ µs, … $t_{11} = 200$ µs. The shutter activation time for the background image was set to $t_b = 650$ µs after the LED on cycle. The shutter on/off times were further tailored to match the LED duty cycle (Figure 2c). Additionally, the camera exposure time for each image was set to match the excitation time (i.e., $t_{excitation} = 70$ ms, also see the *LED driver circuit* section). Camera gain was set to 1. The operation parameters of the PLI reader can be customized to accommodate different lifetime ranges, depending on the emission decay properties of the phosphors.

The *Sweep Delta T* button activated the continuous operation of the PLI, sequentially capturing sets of timelapse phosphorescence images (i.e., 11 timelapse images + 1 background image) with user-specified time intervals in between captures. Furthermore, the GUI contained a screen to display phosphorescence intensity and lifetime image measurements. The GUI had two displaying modes, namely Camera mode and File model. In the Camera mode, the GUI screen displayed real-time images directly captured by the camera, while in the File mode, the displayed images were transferred from a user-specified path. Continuous operation of the reader was set to capture 10 repeated lifetime images within ~45 s time interval and 1 s time gap (Δt) between subsequent images. These images were automatically processed on a benchtop computer to generate final phosphorescence intensity and lifetime images (see the *Generation of phosphorescence intensity and lifetime images* section), which were used in the neural network-based glucose level analysis. The number of phosphorescence intensity images used to generate the final phosphorescence lifetime image was optimized to 6 out of 11, corresponding to optimal linearity, repeatability, and dynamic range of the sensor responses (Figure S12). Total glucose inference time, including image data capture, image processing and deep learning-based analysis, was ~1 min per measurement.

*PLI design*

The PLI consists of the LED driver circuit, camera (UI-3250CP-M-GL Rev.2, IDS) and off-the-shelf optical components arranged within a 3D printed case (~5 cm x 5 cm footprint). The 3D-printed case was printed by the Object 30 printer (Stratasys). The reader contains excitation and emission channels, accommodating spectral properties of the PdBP dyes (630 nm excitation/810 nm emission). The excitation channel contains an LED (633 nm peak wavelength, Osram) mounted to the custom driver circuit, collimating lens (5.0 mm Dia. X 5.0 mm FL, Edmund Optics), excitation filter (632 nm BP, Edmund Optics), and focusing lens (12.0 mm Dia. X 12.0 mm FL, Edmund Optics). The emission channel consists of a camera, imager lens, emission filter (725 nm LP, Edmund Optics), cold mirror (45° AOI, 12.5 mm Square, Edmund Optics), and sample lens (15.0 mm Dia. X 20.0 mm FL, Edmund Optics). The cold mirror directs collimated light from the LED towards the sample plane, generating uniform illumination over a circular area with ~1 cm diameter. Excited phosphorescence emission from the insertable sensor is collected by the emission channel, passing through a two-lens imaging system to generate phosphorescence images of the sensor (Figure 2b). The two-lens system has ~0.75x demagnification, generating 9.6 mm x 7.2 mm FOV. The PLI was designed for a simple mount on the Z-axis stage for *in vitro* testing using glucose-spiked solutions. The cost of an assembled PLI reader was ~$1170, including the camera (see Table S2).

*Generation of phosphorescence intensity and lifetime images*

An automated image analysis algorithm generated the phosphorescence intensity and lifetime images from 10 successive images (Figure 2c). At first, for each of these 10 images, we subtracted the background image from each of the 11 timelapse intensity images, and further performed pixel binning to adjust the image resolution. Next, the pixel-binned images were averaged pixel-wise over the 10 successive measurements, yielding a final set of phosphorescence intensity images. These images were used to construct the phosphorescence lifetime image:

$$S_{n,m}(t_i) = I_{0\,n,m} e^{-t_i/\tau_{n,m}},$$

where $t_i = t_1 \ldots t_{11}$ are the camera shutter activation times for the phosphorescence intensity images (i.e., $t_1 = 0$ μs, …, $t_{11} = 200$ μs), $I_{0\,n,m}$ is the maximal phosphorescence intensity of the pixel located at index $n$, $m$, and $\tau_{n,m}$ is the phosphorescence lifetime of the same pixel. Pixelwise lifetime values at the phosphorescence lifetime image were calculated by solving a least squares optimization problem for $\tau_{n,m}$:

$$\underset{I_{0\,n,m},\,\tau_{n,m}}{\mathrm{argmin}} \; \| \mathbf{log}(\mathbf{S_{n,m}}) - \mathbf{log}(\mathbf{I_{0\,n,m}}) - 1/\tau_{n,m}\mathbf{t} \|,$$

where $\mathbf{log}(\mathbf{S_{n,m}}) = [\log(S_{n,m}(t_1)) \; \log(S_{n,m}(t_2)) \; \ldots \; \log(S_{n,m}(t_{11}))]$ are the pixel-wise phosphorescence intensities, $\mathbf{log}(\mathbf{I_{0\,n,m}}) = \log(I_{0\,n,m})[1 \; \ldots \; 1]_{1 \times 11}$ and $\mathbf{t} = [t_1 \; \ldots \; t_{11}]$.

*In vitro data collection procedures*

The PLI reader and neural-network-based analysis of glucose levels were tested *in vitro* on glucose-spiked DI water samples, utilizing a skin phantom (1 mm, Skin Type 1-2, see the *Skin phantom fabrication* section) to simulate the optical properties of the skin. During *in vitro* testing procedures, a syringe pump drove spiked glucose samples from a stock solution through a fluidic channel containing the sensor (see the *Fluidic channel design and assembly* section) at a constant flow rate of 20 μL/min. Skin phantom was located on top of the fluidic channel, fully covering the sensor (inset on Figure 2a). The fluidic channel was mounted on a custom stage (see the *XY axis stage design* section) for precise alignment between the sensor and the reader. The reader was mounted on a Z-axis optical stage, and the height was adjusted to focus the reader on the sensor plane. The phantom experiments were conducted at room temperature (i.e., 22.8±0.7 °C).

A total of 8 insertable sensors were tested *in vitro* at 10 different glucose concentrations in the 0-250 mg/dL range (i.e., 0, 50, 65, 80, 95, 110, 125, 150, 200, 250 mg/dL). 250 mg/dL glucose sample was prepared by adding 125 mg of glucose (Sigma-Aldrich) into 50 mL DI water. Other glucose samples were prepared by ratiometric mixing of 250 mg/dL sample with DI water. A 15-minute waiting period was used to allow for enzyme activation and signal ramp-up after switching between different glucose levels.

To collect a comprehensive dataset for the neural network-based analysis, each sensor/glucose concentration was tested at 17 discrete locations within the imager FOV, including 5 aligned locations (1 ideal + 4 non-ideal) and 12 misaligned locations arranged into a 4x4 grid (Figure S13). In between subsequent measurements, the custom XY axis stage automatically translated the fluidic channel with the sensor to different locations. The time delay between subsequent data measurements was set to 30 s to allow for the full channel refill with fresh glucose solution. Each data measurement involved 10 phosphorescence lifetime image measurements over ~ 45 s time interval with Δt = 1 s time gap between subsequent measurements (see the *PLI operation* section for more details).

*Fluidic channel design and assembly*

The fluidic channel was assembled from acrylic sheets (1. 5 mm [SOURCEONE] and 1 mm [SimbaLux]), transparency (Apollo) and double-sided tape (3M). The channel shape was precisely cut by a laser cutter (Trotec), and the channel was assembled by stacking 1.5 mm acrylic (bottom part, i.e., floor), 2 transparency sheets, 1 mm acrylic (main part, i.e., channel) and 1 transparency sheet (top part, i.e., ceiling). Double-sided tape was used to glue together the adjacent channel layers. The assembled channel had a size of 75x25x2.8 mm$^3$.

*XY axis stage design*

The XY axis stage was designed to automate sensor misalignments during the *in vitro* testing of the PLI reader. The stage contained 2 stepper motors connected to lead screws, enabling X and Y translation in discrete steps with a 10 µm step size. Stepper motors were controlled by Arduino through stepper motor driver modules. Lead screws were connected to a fluidic channel holder allowing for lateral translation of the channel up to 5 cm along each axis (Figure 2a).

A custom GUI was developed to facilitate the operation of the XY axis stage. The GUI showed the current stage location within the operation range and allowed the user to manually move the stage along the X and Y axes with 2 mm discrete steps. The GUI also enabled the user to align the stage to its origin location (i.e., the default origin location was adjusted to the minimal translation value) and update the origin location with user-controlled values. Finally, the GUI allowed to move the stage based on imported shift values from an external file source. Stage operation code and GUI were developed in Python, using the Tkinter package to build the GUI and the Serial package to communicate with Arduino. For the automated misalignment data collection with the PLI reader, the stage operation code was integrated into the PLI reader operation code.

*Neural network-based analysis of PLI data*

Neural network-based sensor analysis was implemented to (1) assess proper PLI reader alignment and (2) classify glucose concentrations between low, medium, and high ranges for the properly aligned samples. The neural network-based analysis consisted of two separate CNNs for the alignment (CNN$_{Alignment}$) and the classification (CNN$_{Class}$) tasks (Figure 4a). The alignment network was trained using phosphorescence intensity images, while the classification network was trained utilizing phosphorescence lifetime images. We did not use phosphorescence intensity images to train the classification model due to the high inter-sensor variability of phosphorescence intensity signals. Prior to input into the CNN$_{Class}$, the phosphorescence lifetime images were cropped to center the sensor location within the image and standardized by the maximal lifetime value (i.e., 250 µs); higher lifetime values at phosphorescence lifetime images were set to 250 µs to reduce background noise.

CNN$_{Alignment}$ contained 2 convolutional layers (32 units with 3x3 kernel and 16 units with 2x2 kernel) followed by a flattening layer and two fully-connected layers (128 and 32 units). Both convolutional and fully-connected layers had 'ReLU' activation functions and L2 regularization. The output layer had 1 unit with a 'sigmoid' activation function. The network used a binary cross-entropy loss complied with the Adam optimizer, a learning rate of e$^{-3}$ and a batch size of 5. Binary cross-entropy loss ($L_{BCE}$) is defined as:

$$L_{BCE}(y, y') = -\frac{1}{N_b}\Sigma_{i=1}^{N_b}\bigl(y_i \log(y'_i) - (1 - y_i)\log(1 - y'_i)\bigr),$$

where $y_i$ are the ground truth labels (i.e., "1" for samples located within the aligned region, and "0" for samples located within the misaligned region), $y'_i$ are the predicted labels and $N_b$ is the batch size.

CNN$_{Class}$ consisted of 3 convolutional layers (512 units with 5x5 kernel, 256 units with 3x3 kernel and 128 units with 3x3 kernel) followed by a flattening layer and two fully-connected layers (128 and 32 units). All layers had 'ReLU' activation functions and used L2 regularization. The output layer had 3 units with 'sigmoid' activation functions. The loss function for this network was the categorical cross-entropy loss compiled with Adam optimizer, a learning rate of e$^{-4}$ and a batch size of 5. Categorical cross-entropy loss ($L_{CCE}$) is defined as:

$$L_{CCE}(y, y') = -\frac{1}{N_b}\sum_{i=1}^{N_b}\sum_{p=1}^{3}\left(y_{i,p}\log\left(y'_{i,p}\right)\right),$$

where p = 1, 2, 3 is the glucose level class (i.e., 1 – *low*, 2 – *normal*, and 3 – *high*), $y_{i,p}$ are the ground truth classification labels, $y'_{i,p}$ are the predicted classification labels and $N_b$ is the batch size.

The alignment neural network was optimized and trained on a training/validation set with a total of 680 measurements, including 4 different sensors, 10 glucose concentrations per sensor and 17 locations per concentration. CNN$_{Alignment}$ achieved 100% accuracy on the validation data through 4-fold cross-validation, correctly identifying all the aligned and misaligned locations. The network was further blindly tested on a testing set composed of 672 samples (i.e., 4 sensors x 10 glucose concentrations x 17 locations; the number is <680 due to the unexpected experiment interruptions/failures for a few glucose concentrations) and achieved a blind testing accuracy of 100%. More results about the performance of the CNN$_{Alignment}$ model on the testing set are reported in the *Neural network-based analysis of PLI data for inferring glucose concentration levels* section.

CNN$_{Class}$ was solely trained on sensors located within the alignment region, and the total number of training/validation samples was 200 (i.e., 4 sensors x 10 glucose concentrations x 5 *aligned* locations). CNN$_{Class}$ achieved 89.3% accuracy on the validation data, including an accuracy of 93.7% and 87.5% for the ideal and non-ideal locations, respectively. The blind testing set included 198 samples, and CNN$_{Class}$ achieved 88.8% blind testing accuracy, including accuracies of 92.5% and 87.3% for the ideal and non-ideal locations, respectively. More details about CNN$_{Class}$ performance on the blind testing set are reported in the *Neural network-based analysis of PLI data for inferring glucose concentration levels* section.

Training times for CNN$_{Alignment}$ and CNN$_{Class}$ were 5 min and 30 min respectively. Blind testing times of trained CNN$_{Alignment}$ and CNN$_{Class}$ models we considerably lower, with ~20 ms per measurement. Data preprocessing to generate phosphorescence intensity and lifetime image sets was done in MATLAB 2023b, and the training/testing of the neural networks was performed in Python, using OpenCV and TensorFlow libraries. Training/testing of the neural networks was done on a desktop computer with a GeForce GT 1080 Ti (NVIDIA).

*Skin phantom fabrication*

To create tissue-simulating optical phantoms, PDMS (Sylgard™ 184 Silicon Elastomer Kit) was used as the base, and TiO$_2$ and carbon black (Carbon, mesoporous nanopowder, Millipore Sigma) were added as the scattering and absorbing agents, respectively. To fabricate a 1-mm thick phantom mimicking lighter skin type (i.e., Type 1-2, Fitzpatrick scale has an absorption coefficient [$\mu_a$] of 0.04-0.14 cm$^{-1}$ and a reduced scattering coefficient [$\mu'_s$] of 10-15 cm$^{-1}$ [66]), 2 grams of PDMS were weighed onto a 2-inch diameter aluminum mold. A 0.0025% (w/w) carbon black to PDMS and 0.05% (w/w) of TiO$_2$ to PDMS were then added to 0.2 grams of curing agent in a separate glass beaker. This beaker was sonicated for 30 min to break up $TiO_2$ powder clumps and mixed using a glass rod to create a homogeneous mixture. The $TiO_2$ and carbon black mixture was then transferred to the PDMS in the aluminum mold and mixed thoroughly using a glass rod to ensure a homogenous mixture. Next, a vacuum chamber was used to degas the mixture in the mold for 30 min and remove any air bubbles trapped within the viscous media. The mold was then placed in a toaster oven and heated at 150 ºC for 4 min to get a completely cured phantom. The silicon phantom was carefully removed from the mold using a lab spatula. The finished disc phantoms were optically characterized using a laser source at 633 nm, a single integrating sphere, and an inverse adding-doubling algorithm, yielding an absorption coefficient of 0.05 $cm^{-1}$ and a reduced scattering coefficient of 13.9 $cm^{-1}$ with anisotropy, g = 0.9. A rectangular slab was then cut out of this disc phantom for the *in vitro* testing of the insertable glucose sensors (Figure 2a)[67].

## Supporting Information Available

Supplementary Tables 1-2; Supplementary Figures S1-S10; Supplementary Video 1

[66] Karsten, A. E.; Singh, A.; Karsten, P. A.; Braun, M. W. Diffuse reflectance spectroscopy as a tool to measure the absorption coefficient in skin: South African skin phototypes. *Photochemistry and photobiology* **2013**, 89(1), 227-233, DOI: 10.1111/j.1751-1097.2012.01220.x;

[67] Prahl, S. Everything I think you should know about Inverse Adding-Doubling. *Oregon Medical Laser Center, St. Vincent Hospital* **2011**, 1344, 1-74.20

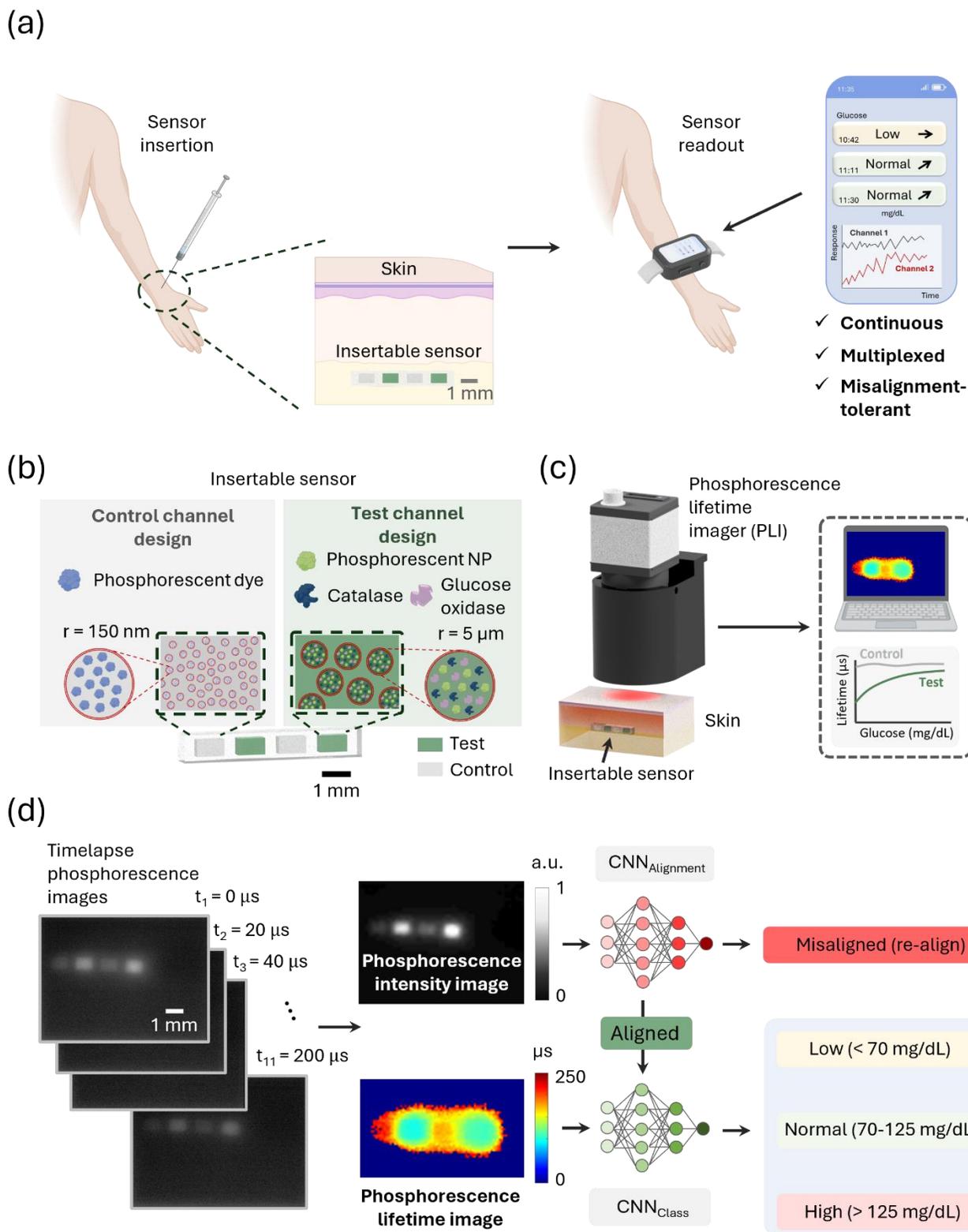

**Figure 1**. (a) Overview of optical continuous glucose monitoring, featuring an insertable/implantable passive biosensor and a wearable reader; (b) Insertable sensor structure; (c) Operation of the phosphorescence lifetime imager (PLI); (d) Neural network-based processing of phosphorescence images for misalignment-resilient classification of glucose concentration levels (low/normal/high).
21

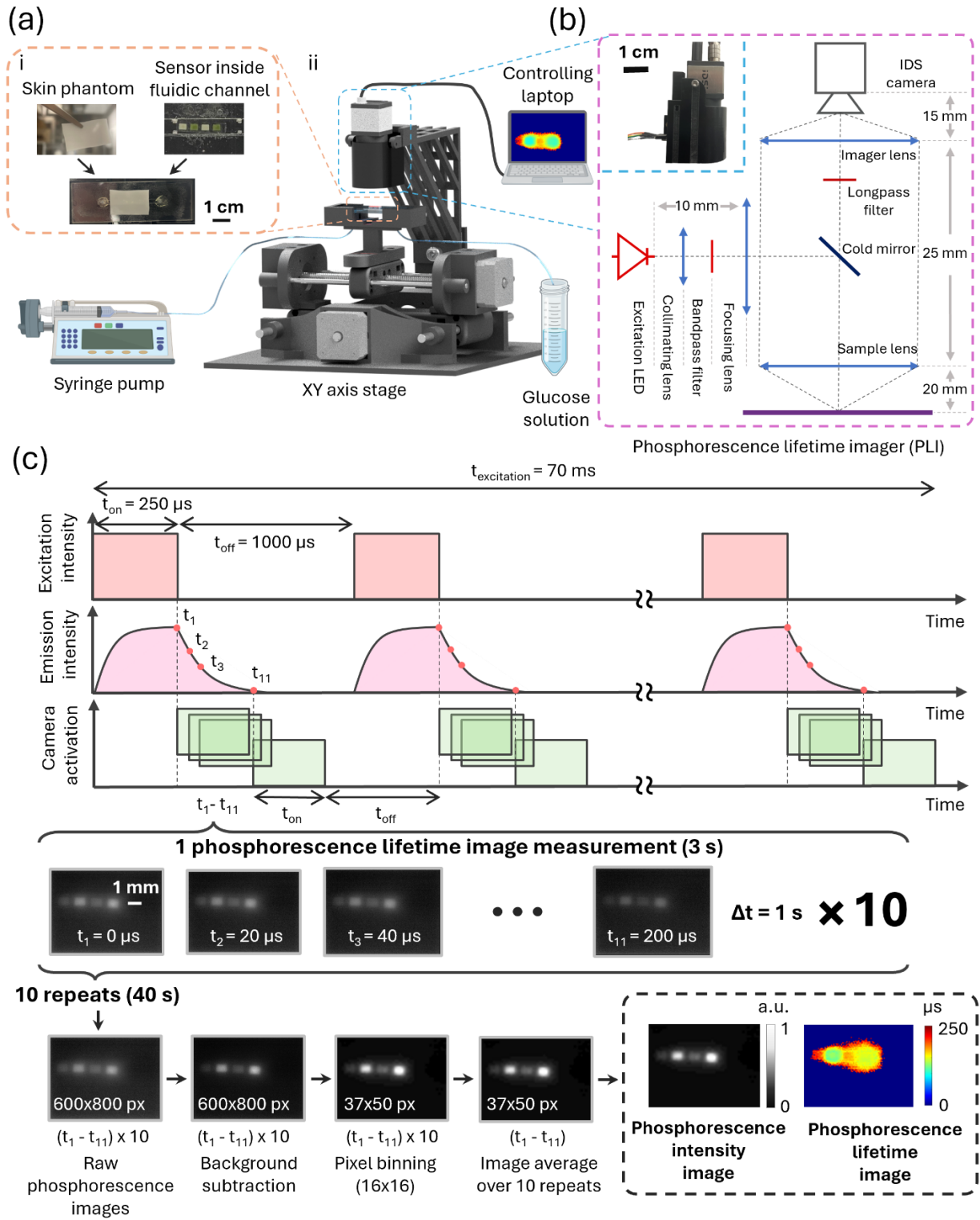

**Figure 2**. (a) Experimental setup for *in vitro* testing of the PLI reader; (b) PLI reader structure; (c) Reconstruction of the phosphorescence intensity and phosphorescence lifetime images.

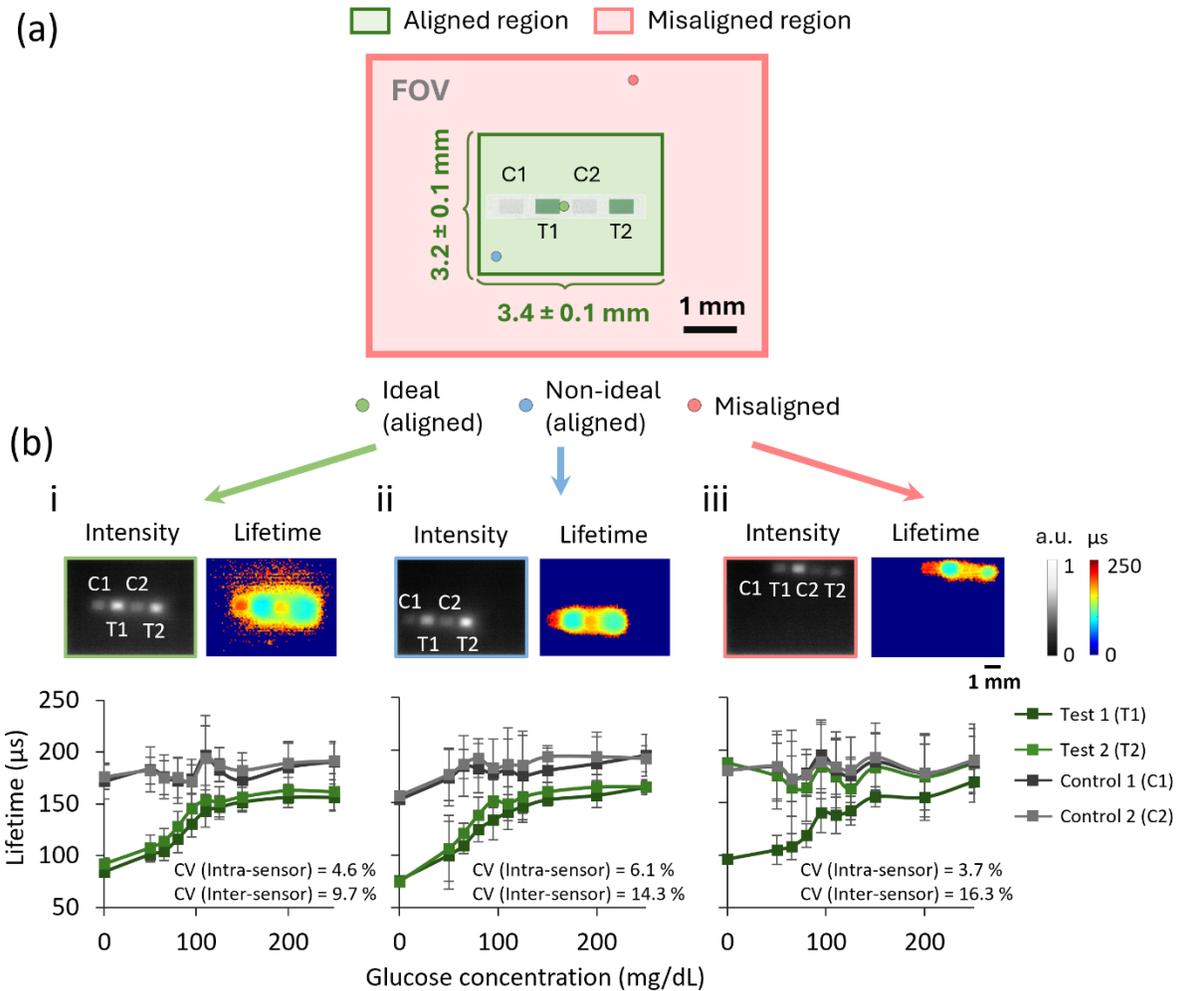

**Figure 3**. (a) PLI reader field-of-view (FOV) with *aligned* and *misaligned* regions. Aligned region, represented by a 3.2 mm x 3.4 mm rectangle, is a misalignment-tolerant zone. When the implanted sensor location is within the aligned region, the PLI reader and the trained neural network accurately perform glucose level inference despite random reader misalignments from the ideal location (i.e., FOV center). If the implanted sensor falls into the misalignment region, the reader will prompt the user for re-alignment. (b) Lifetime responses of the insertable sensors for 0-250 mg/dL glucose concentration range captured by the PLI reader for three representative sensor locations withing the FOV, including (i) the ideal location (i.e., the center of the aligned region), (ii) a non-ideal location (i.e., the border of the aligned region), and (iii) a misaligned location within the misaligned region.

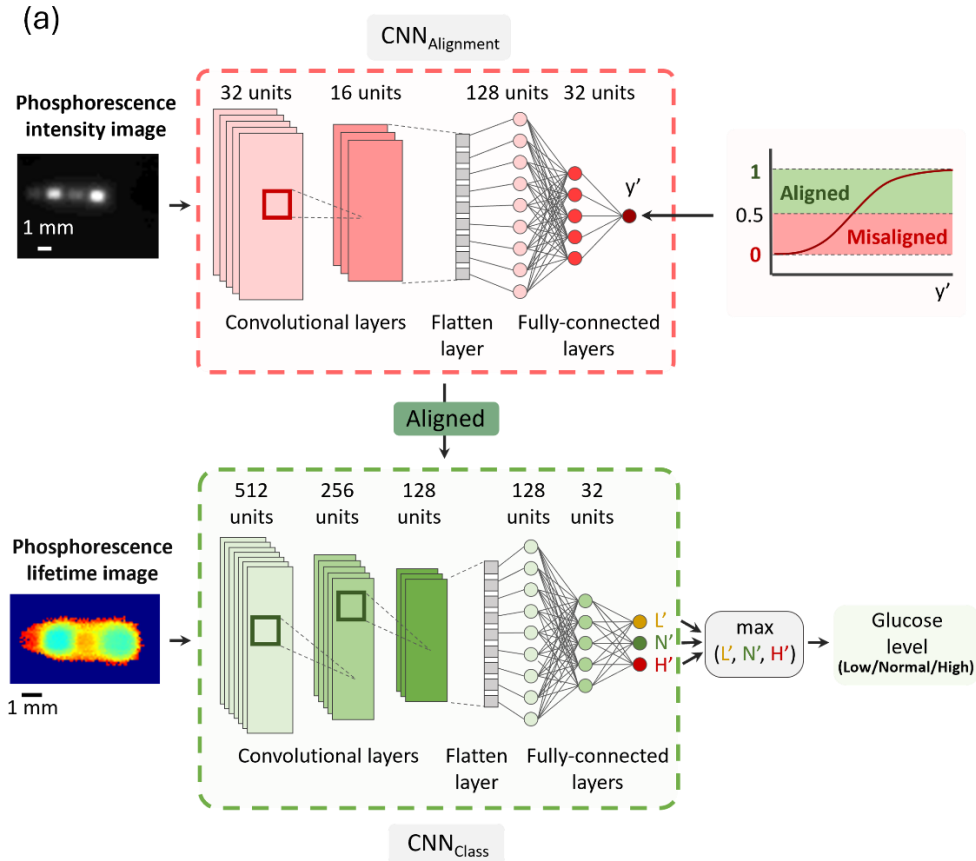

**Figure 4**. (a) Neural network-based misalignment-resilient glucose level inference framework. The framework consists of two separate convolutional neural networks, namely the alignment network (CNN$_{Alignment}$) and the classification network (CNN$_{Class}$). First, phosphorescence intensity images acquired by the PLI are processed by CNN$_{Alignment}$, which classifies the reader location as *aligned* or *misaligned*. If the PLI reader position is identified as *misaligned*, the reader prompts the user for re-alignment and CNN$_{Class}$ is not used in this case. If the reader location is identified as *aligned*, the PLI image data are further processed by the second neural network, CNN$_{Class}$, which classifies the glucose concentration levels (low/normal/high) using phosphorescence lifetime measurements as input data; (b) Blind testing results for CNN$_{Alignment}$; (c) Blind testing results for CNN$_{Class}$.

24